\documentclass[twocolumn,superscriptaddress,nofootinbib,floatfix,aps,showpacs,prb,citeautoscript,reprint]{revtex4-1}
\usepackage{dcolumn}
\usepackage{bm}
\usepackage{amsmath}
\usepackage{amsfonts}
\usepackage{amssymb}
\usepackage{graphicx}
\usepackage{tabularx}
\usepackage{subfigure}
\usepackage{color}
\usepackage{gensymb}
\usepackage{pgfgantt,rotating}
\usepackage{physics}
\usepackage{array,multirow}
\usepackage{threeparttable}
\usepackage{tablefootnote}
\usepackage{adjustbox}
\usepackage{mathtools}
\usepackage{enumitem}
\usepackage{wasysym}
\usepackage{framed}
\usepackage{subfloat}
\usepackage{blindtext}

\usepackage{hyperref} \hypersetup{colorlinks=true, linkcolor=blue,
  citecolor=blue, urlcolor=blue, pdftitle={First-principles investigation of graphene/MoS$_2$ bilayer heterostructures using Tkatchenko-Scheffler van der Waals method}}

\begin{document}

\title{First-principles investigation of graphene/MoS$_2$ bilayer heterostructures using Tkatchenko-Scheffler van der Waals method}

\author{Sobhit Singh}
\email{smsingh@mix.wvu.edu}
\affiliation{Department of Physics and Astronomy, West Virginia University, Morgantown, WV-26505-6315, USA}

\author{Camilo Espejo}
\affiliation{Departamento de Ciencias B\'asicas y Modelado, Universidad Jorge Tadeo Lozano, Cra 4 No 22-61, Bogot\'a D.C, Colombia}

\author{Aldo H. Romero}
\affiliation{Department of Physics and Astronomy, West Virginia University, Morgantown, WV-26505-6315, USA}


\begin{abstract}
Graphene/MoS$_2$ van der Waals (vdW) heterostructures have promising technological applications due to their unique properties and functionalities. Many experimental and theoretical research groups across the globe have made outstanding contributions to benchmark the properties of graphene/MoS$_2$ heterostructures. Even though some research groups have already made an attempt to model the graphene/MoS$_2$ heterostructures using {\it first-principles} calculations, there exists several discrepancies in the results from different theoretical research groups and the experimental findings. In the present work, we revisit this problem by first principles approach and address the existing discrepancies about the interlayer spacing between graphene and MoS$_2$ monolayers in graphene/MoS$_2$ heterostructures, and the location of Dirac points near Fermi-level. We find that the Tkatchenko--Scheffler method efficiently evaluates the long-range vdW interactions and accurately predicts interlayer spacing between graphene and MoS$_2$ sheets. We further investigate the electronic, mechanical and vibrational properties of the optimized graphene/MoS$_2$ heterostructures created using 5$\times$5/4$\times$4 and 4$\times$4/3$\times$3 supercell geometries having different magnitudes of lattice mismatch. The effect of the varying interlayer spacing on the electronic properties of heterostructures is discussed. Our phonon calculations reveal that the interlayer shear and breathing phonon modes, which are very sensitive to the weak vdW interactions, play vital role in describing the thermal properties of the studied systems. The thermodynamic and elastic properties of heterostructures are further discussed. A comparison between our results and the results reported from other research groups is presented. 
\end{abstract}


\maketitle
\section{Introduction}
The proximity of a substrate material is known to significantly change the electronic and optical properties of the widely celebrated graphene \cite{Novoselov2014, Novoselov2012, Castro_Rev2009, WeiHan2014, Novoselov2016, sobhitProximity2018}. Recently, this particular issue has been comprehensively investigated, both theoretically and experimentally, due to the promising applications of graphene-based van der Waals (vdW) heterostructures in the modern spintronics and optoelectronics industry \cite{Xiaofeng2015, Novoselov2016}. The advanced material fabrication techniques have enabled us to stack different layers of materials in a controlled manner, and fabricate the desired vdW heterostructures for targeted applications. Graphene/MoS$_2$ heterostructures are one of the most prominent vdW heterostructures that have been successfully synthesized in laboratory \cite{RadisavljevicB_Nature2011, RoyNature2013, SimoneACSNano2013, SupNature2013, Britnell_Science2013, Larentis_Nano2014, LiliACSNano2014}, and are proven to exhibit intriguing physical and chemical properties. \cite{Chhowalla_NatureChem2013, LiliACSNano2014}  The presence of MoS$_2$ substrate induces strong spin-orbit coupling (SOC) of strength $\sim$1 meV in graphene, which is almost 1000 times larger than the intrinsic SOC of pristine graphene, and consequently it opens a bandgap at Dirac point in graphene. \cite{LuPRL2014} Recent studies report the observation of exceptional optical response with large quantum efficiency, gate-tunable persistent photoconductivity, photocurrent generation, and negative compressibility in graphene/MoS$_2$ heterostructures. \cite{RoyNature2013, SimoneACSNano2013, SupNature2013, Britnell_Science2013, Larentis_Nano2014, LiliACSNano2014}  Electronic logic gates, memory devices, optical switches, energy conversion and storage devices,  catalysts, and nanosensors have already been constructed using graphene/MoS$_2$ heterostructures.\cite{RoyNature2013, SimoneACSNano2013, SupNature2013, Britnell_Science2013, Zhang_gMoS2-2014, Larentis_Nano2014, Novoselov2016, Deng2016} Furthermore, graphene/MoS$_2$ heterostructures intercalated with selected metals have been extensively investigated due to their extraordinary energy storage capacity and novel chemical properties. \cite{Chang2011, ChangCC2011, ShaoJPCC2015, ZeljkoPRM2017}

Large lattice mismatch and presence of weak vdW interaction between graphene and MoS$_2$ monolayers make the Density Functional Theory (DFT)\cite{HK1964, KS1965} modeling of graphene/MoS$_2$ heterostructures computationally challenging. Although the lattice mismatch can be minimized by stacking commensurate supercells of graphene and MoS$_2$ monolayers, the correct {\it first-principles} determination of weak non-local vdW interactions remains elusive in graphene/MoS$_2$ heterostructures. Various different methods have been employed to predict the correct interlayer spacing between graphene and MoS$_2$ monolayers. Using the semiempirical DFT-D2 method of Grimme \cite{Grimme2006} and 4$\times$4/3$\times$3 (hereafter 4:3) supercell geometry of graphene/MoS$_2$, Gmitra et al.  \cite{GmitraPRB2015} have predicted the interlayer spacing of 3.37 \AA. They also observed that the Dirac point of graphene is located close to the conduction band of MoS$_2$ indicating the enhanced screening and substantial increase in the mean free path of carriers in the graphene layer. \cite{GmitraPRB2015} Moreover, for the first time Gmitra et al. \cite{GmitraPRB2015} theoretically demonstrated the use of graphene/MoS$_2$ heterostructures as a platform for optospintronic devices. Ebnonnasir et al. \cite{Abbas_APL2014} studied two commensurate graphene/MoS$_2$ bilayer heterostructures constructed using 5$\times$5/4$\times$4 (hereafter 5:4) and 4$\times$4/3$\sqrt{3}$$\times$3$\sqrt{3}$ supercells of graphene/MoS$_2$. Using the vdW exchange-correlation functional of Klime{\v{s}} et al. \cite{Klime2011} (optB86b-vdW), they predict the interlayer spacing of 3.11 and 3.13 \AA~for 5:4 and 4:3 heterostructure systems, respectively. Contrary to the report of Gmitra et al.\cite{GmitraPRB2015}, the Dirac point in ref. \cite{Abbas_APL2014} lies at the Fermi-level for 5:4 geometry and slightly above the Fermi-level for 4:3 geometry.  Shao et al. \cite{ShaoJPCC2015} reported an interlayer spacing of 3.37 \AA~for graphene/MoS$_2$ 4:3 bilayer system treated with Klime{\v{s}} et al. \cite{Klime2011} vdW functional. They found that the Dirac point is located near the conduction band of MoS$_2$ yet it is within the energy bandgap region. Li et al. \cite{XDLi2013} reported an interlayer spacing of 3.36 \AA~ for 4:3 graphene/MoS$_2$ bilayer, they also report that the Dirac point is located above the Fermi-level touching the conduction band of MoS$_2$ and therefore, indicating charge transfer between graphene and MoS$_2$ layers as reported by Gmitra et al. \cite{GmitraPRB2015} 

Using different {\it ab-initio} codes and different implementation for vdW corrections, several theoretically studies predicted graphene/MoS$_2$ interlayer spacing ranging from 3.11 to 4.32 \AA~[see supplemental information (SI)]~\cite{Yandongma2011graphene, Abbas_APL2014, ShaoJPCC2015, GmitraPRB2015, NamLe2016, hu2016effects, ZeljkoPRM2017, NGUYEN201879}. Also, different works inconsistently report distinct location of the Dirac point near Fermi-level and incoherently argue about the charge transfer mechanism in this system. However, the experimental investigations on graphene/MoS$_2$ bilayer systems reveal an interlayer spacing of 3.40 $\pm$ 0.1 \AA~and suggest no charge transfer between the layers at equilibrium conditions.\cite{PierucciNano2016} Furthermore, although the vibrational, elastic and mechanical properties of pristine graphene and MoS$_2$ monolayers have been thoroughly studied, \cite{Yan2008, Sofo2007, Wang2009, Lui2013,  DING2011, Zhao2013, Jiang2015, ToheiPRB2006, YUAN20151, Lee385, Cooper2013, Jiang_review2015, AndrewPRB_2012, Xiaoding2009, KaiNano2014,Qing_MoS2-2013} a little attention has been paid to the aforementioned properties of their heterostructures.\cite{NamLe2016, KaiNano2014}

In this work, we revisit the problem of accurate evaluation of weak non-local vdW interactions in graphene/MoS$_2$ bilayer heterostructures using different methodologies. We find that the DFT-D2~\cite{Grimme2006} and Tkatchenko--Scheffler (DFT-TS)~\cite{TS_PRL2009} methods yield very good estimate of the interlayer spacing in graphene/MoS$_2$ bilayer heterostructure. We further present a detailed characterization of the electronic and vibrational (phonons) properties of the graphene/MoS$_2$ heterostructures using DFT-TS method. We have studied the two most commonly used 5:4 and 4:3 supercell geometries of graphene/MoS$_2$ heterostructures. To resolve the noted discrepancy about the location of Dirac point in the energy space, the effect of varying interlayer spacing on the electronic properties of graphene/MoS$_2$ heterostructures is investigated. Our results indicate that the heat capacity and overall elastic properties of graphene/MoS$_2$ heterostructures are considerably better compared to that of the individual monolayers.

\section{Computational Details}
Density Functional Theory (DFT) \cite{HK1964, KS1965} based first-principles calculations were carried out using the projector augmented-wave (PAW) method as implemented in the {\sc VASP} code \cite{Kresse1996, Kresse1999}, using the Perdew-Burke-Ernzerhof (PBE) parametrized generalized gradient approximation (GGA) exchange-correlation functional.\cite{Perdew1996} The SOC was employed by a second-variation method implemented in the {\sc VASP} code. We considered twelve valence electrons of Mo (4$p^6$, 5$s^1$, 4$d^5$), six valence electrons of S (3$s^2$, 3$p^4$), and four valence electrons of C (2$s^2$, 2$p^2$) in the PAW pseudo-potential. We consider two commensurate supercell geometries: (i) 5:4 and (ii) 4:3, to minimize the lattice mismatch between graphene and MoS$_2$ layers. A vacuum of thickness larger than 17 \AA~was added along $c$-axis to avoid the periodic interactions. The lattice parameters and the inner coordinates of atoms were optimized until the Hellmann-Feynman residual forces were less than $10^{-4}$ eV/{\AA} per atom, and $10^{-8}$ eV was defined as the total energy difference criterion for convergence of electronic self-consistent calculations. SOC and vdW interactions (DFT-TS) \cite{TS_PRL2009, Bu_TSvdW2013} were included in the structural optimization. We used $650$ eV as the kinetic energy cutoff of plane wave basis set and a $\Gamma$-type $10\cross10\cross1$ $k$-point mesh was employed to sample the irreducible Brillouin zone of heterostructures. The phonon calculations were performed using density functional perturbation theory ({\sc DFPT}) approach, and {\sc PHONOPY} code \cite{Togo2008} was used for the post-processing of data. To investigate the effect of uniaxial stress along $c$-axis, we varied the interlayer spacing from $-4\%$ (compression) to $+4\%$ (expansion). The inner coordinates of all atoms in the strained cell were relaxed only along $x-y$ directions while maintaining their $z$ coordinates frozen. The {\sc PyProcar} code\cite{PyProcar, sobhit2016PRB, singh2016PCCP} was used to plot the spin-projected electronic bands, and the {\sc MechElastic} script\cite{MechElastic} was used to evaluate the elastic properties of graphene/MoS$_2$ bilayer heterostructures.

Complementary calculations were performed using {\sc ABINIT}\cite{GONZE2016106} code in order to guarantee that some of the observed physical properties are independent of the details of the numerical implementation and the approximations made to account for core electrons. In particular, the interlayer distance for 5:4 bilayer heterostructure and its corresponding band structure were evaluated. In order to calculate the interlayer equilibrium distance we performed DFT-D3 calculations\cite{Grimme2010} omitting three-body contributions to the vdW energy \cite{dft-dinabinit}. Terms involving two-atoms were taken into account for vdW correction if their energy contribution was larger than $10^{-9}$ eV. The exchange-correlation functional used in {\sc ABINIT} was chosen to be same as the one used in {\sc VASP} calculations, however the criterion for convergence of the electronic self-consistent calculations was $10^{-9}$ eV, and the plane wave cutoff was 980 eV. Optimized norm-conserving Vanderbilt pseudopotentials \cite{Hamann2013} from the PseudoDojo \cite{pseudodojo} were used. The number of valence electrons for S and C atoms was same as  mentioned earlier, while for Mo we considered 14 electrons (4$s^6$, 4$p^6$, 5$s^1$, 4$d^5$). No SOC was considered in the {\sc ABINIT} calculations.


\begin{figure}[hb!]
 \centering
 \includegraphics[width=8.7cm, keepaspectratio=true]{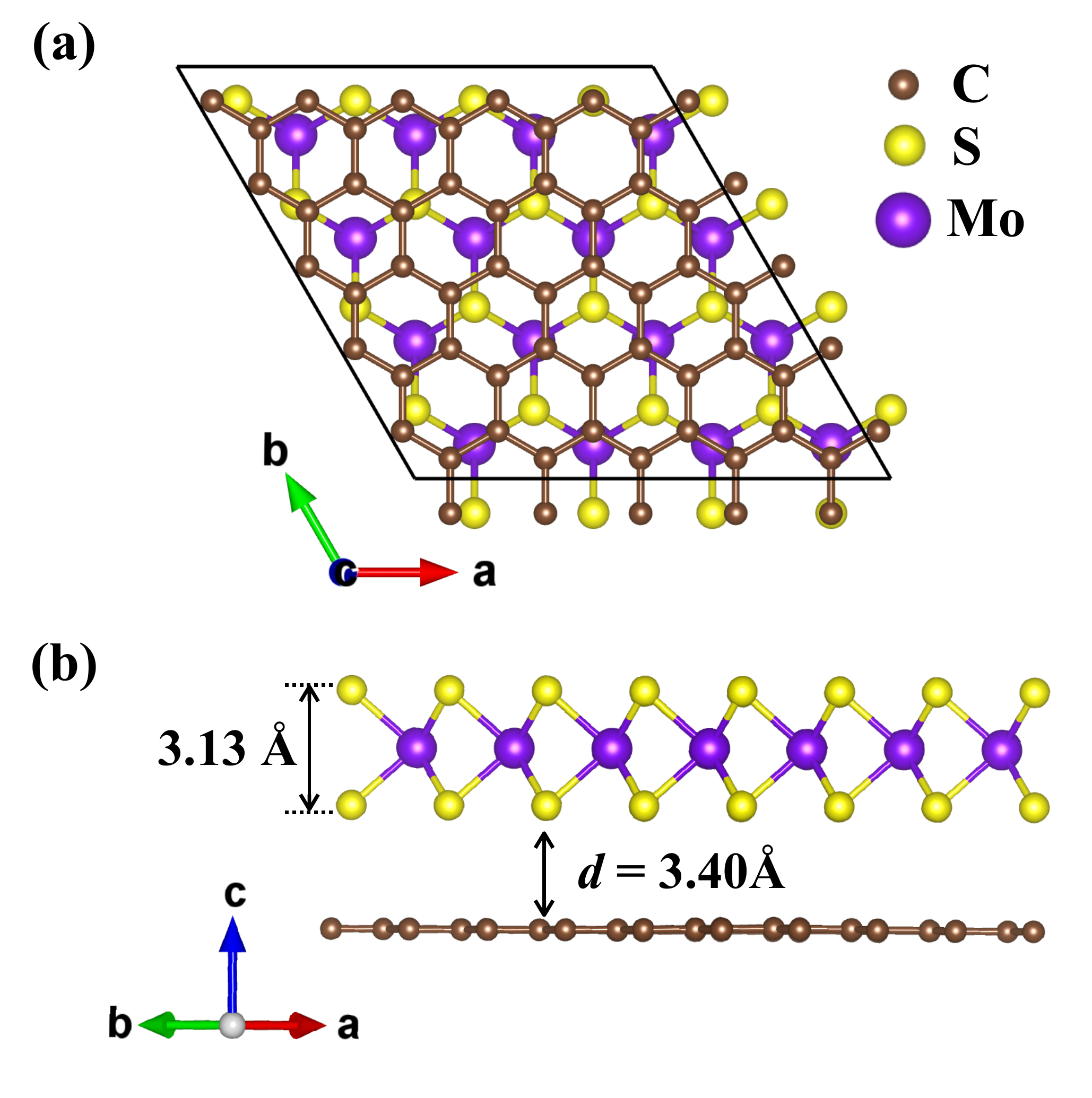}
 \caption{(Color online) Figures (a-b) represent the crystal structure of 5:4 graphene/MoS$_2$ bilayer heterostructure from two different crystal orientations. }
 \label{fig:mono_crystal-phonons}
 \end{figure}

\section{Results and discussion}

In the following sections, we present our results addressing the structural, electronic, vibrational, and elastic properties of graphene/MoS$_2$ bilayers. 

\subsection{Crystal structure of graphene/MoS$_2$ vdW heterostructures: Evaluation of the accurate vdW interaction}

Figure \ref{fig:mono_crystal-phonons} shows the crystal structure of 5:4 graphene/MoS$_2$ bilayer heterostructure from two different crystal orientations. Since the 5:4 supercell geometry has lower lattice mismatch compared to that of the 4:3 combination, we decide to discuss the optimized crystal details for the 5:4 geometry here, however, details of the 4:3 structure are given in Table 1. The lattice parameters of SOC+vdW optimized 5:4 bilayer heterostructure are $a = b = $12.443 \AA. In this case, MoS$_2$ sheet is being compressed by 0.3\%, whereas the graphene sheet is being stretched by 1.16\% from the optimized pristine cell parameters. Mo-S bond length is 2.38 \AA~and C-C bond length is 1.44 \AA. The thickness of the MoS$_2$ monolayer, i.e. the vertical distance between S-S planes, is 3.13 \AA. The interlayer spacing ($d$) between graphene and MoS$_2$ sheets is 3.40 \AA, which is in excellent agreement with the experimentally reported interlayer spacing of 3.40 \AA. \cite{PierucciNano2016} However, the interlayer spacing ($d$) for 4:3 supercell geometry is 3.42 \AA, which is slightly larger than the case of 5:4 geometry. This can be ascribed to the relatively large lattice mismatch between graphene and MoS$_2$ monolayers in 4:3 geometry. On the other hand, DFT-D3 structural optimization in {\sc ABINIT} yields very similar {\it intralayer} bond lenghts, while it underestimates the interlayer spacing. The latter was found to be 3.3 \AA. Therefore, it is worth to note that the DFT-TS method \cite{TS_PRL2009, Bu_TSvdW2013} remarkably describes the weak vdW interactions accurately in the graphene/MoS$_2$ bilayers, whereas other semiempirical vdW methods \cite{Grimme2006, Grimme2010, Klime2011, LEBEDEVA201745} appear to be not as accurate in describing this system. 

The main reason behind the success of the DFT-TS method compared to the other vdW methods is the fact that the studied heterostructures consist of semimetallic and semi-insulating monolayers, thus having a large variation in the local charge density distribution at the interface (see Fig.~\ref{isosurf}). Formally, the TS-method is similar to the semiempirical DFT-D2 method of Grimme \cite{Grimme2006}. The key difference between these methods is that the vdW parameters, dispersion coefficients ($\displaystyle C_{6ij}$) and the damping function $\displaystyle f(r_{ij})$ that scale the force field in vdW systems, are fixed for each element in DFT-D2 method and thus, these parameters are insensitive to the chemical environment of the system. Here, $i$ and $j$ refer to the atomic indices. For instance, the $C{_6}$ parameter for carbon atom in methane, diamond and graphene takes exactly same values. On the other hand, in the DFT-TS method vdW parameters, $\displaystyle C_{6ij}$ and $\displaystyle f(r_{ij})$ are charge-density dependent. Therefore, the DFT-TS method is sensitive to the chemical environment. Since the DFT-TS method relies on the summation of the $\displaystyle C_{6}$ coefficients, derived from the electronic density of atoms in solid, and accurate reference data for free atom, the DFT-TS method efficiently accounts variations in vdW contribution of atoms due to their local chemical environment. In this method, polarizability ($\alpha_i$), $\displaystyle C_{6ij}$ and atomic radii of atom $R_{0i}$ are calculated using the following relations:

\begin{equation}
   {\alpha{_{i}} = \nu{_{i}} \alpha{_{i}}^{free}} 
 \label{eqn:1}
\end{equation}

\begin{equation}
  {C_{6ii} = \nu{_{i}^2} C_{6ii}^{free}}
  \label{eqn:2}
\end{equation}

\begin{equation}
    {R_{0i} = \left(\frac{\alpha_{i}}{\alpha_{i}^{free}} \right)^{\frac{1}{3}} R_{0i}^{free}}
  \label{eqn:3}
\end{equation}

The $\alpha_{i}^{free}$, $C_{6ii}^{free}$ and $R_{0i}^{free}$ parameters for free atoms are taken from the reference data and effective atomic volumes $ \nu_{i}$ are determined using the Hirshfeld partitioning of the all-electron density: \cite{Hirshfeld1971}

\begin{equation}
     { \nu_{i} = \frac{\int r^3 \,w_i({\mathbf{r}}) n({\mathbf{r}})\, d^3{\mathbf{r}}}{\int r^3\, n_{i}^{free}({\mathbf{r}})\,d^3{\mathbf{r}}}}
  \label{eqn:4}
\end{equation}

Here,  $n({\mathbf{r}})$ is the total electron density,  $ n_{i}^{free}({\mathbf{r}})$ is the spherically averaged electron density of the neutral free atomic species $i$, and $ w_i({\mathbf{r}})$ is the Hirshfeld weight.\cite{Hirshfeld1971}

The optimized vdW parameters for graphene/MoS$_2$ bilayer heterostructures are listed in Table 1.  The cutoff radius for pairing interactions was set to 30 \AA~ and the damping parameters and the reference data for free atomic species were used as implemented in the {\sc VASP} code. For more technical details of the used methodology and application of the TS-method for dispersion corrections to DFT energies and forces to extended systems including noble-gas solids, molecular crystals, layered and chain-like structures, ionic crystals, and metals, we refer the reader to the excellent paper by Bu{\v{c}}ko et al.\cite{Bu_TSvdW2013}

\newcolumntype{L}{>{\centering\arraybackslash}m{3cm}}
\begin{table*}[htb!]
\centering
\begin{center}
\caption{The parameters for pair-potential, namely-- dispersion coefficient $C_{6}$ , atomic radius $R_{0}$ and polarizability $\alpha$ (all in a.u.), obtained from the TS-method, optimized interlayer spacing $d$ (\AA), bond length (\AA), and optimized lattice parameters (\AA) for 2D graphene/MoS$_2$ bilayer heterostructures are listed. \\ }

\begin{adjustbox}{width=1.0\textwidth, center=\textwidth } 
   \begin{threeparttable}
   \setlength{\arrayrulewidth}{0.3mm}
\renewcommand{\arraystretch}{1}
\begin{tabular}{L  L   L  L  m{1cm}   c  c}
\hline
System & $C_{6}$ & $R_{0}$ & $\alpha$ & $d$  & bond length & lattice parameters  \\
\hline\\ 
\multirow{3}{3cm}{5:4 heterostructure}  &  Mo: 1040.3  &  Mo: 4.1  & Mo: 88.9  &  \multirow{3}{*}{3.40}  &  Mo-S bond = 2.38   &  \multirow{3}{*}{$a$ = $b$ = 12.443}  \\
&  S: 129.3  &  S: 3.8  & S: 19.3  &   &  C-C bond = 1.44  &   \\
&  C: 35  &  C: 3.4  & C: 10.4  &   &  S-S bond = 3.13   & \\ \\
\hline\\ 
\multirow{3}{3cm}{4:3 heterostructure}  &  Mo: 1036.7  &  Mo: 4.1  & Mo: 88.8  &  \multirow{3}{*}{3.42}  &  Mo-S bond = 2.42   &  \multirow{3}{*}{$a$ = $b$ = 9.698}  \\
&  S: 129.0  &  S: 3.8  & S: 19.2  &   &  C-C bond = 1.40  &  \\
&  C: 34.2  &  C: 3.4  & C: 10.3  &   &  S-S bond = 3.08  &   \\ \\

\hline
\end{tabular}
 \end{threeparttable} 
\end{adjustbox}
\label{table1}
\end{center}
\end{table*}

\begin{figure*}[hbt!]
 \centering
 \includegraphics[width=18.0cm, keepaspectratio=true]{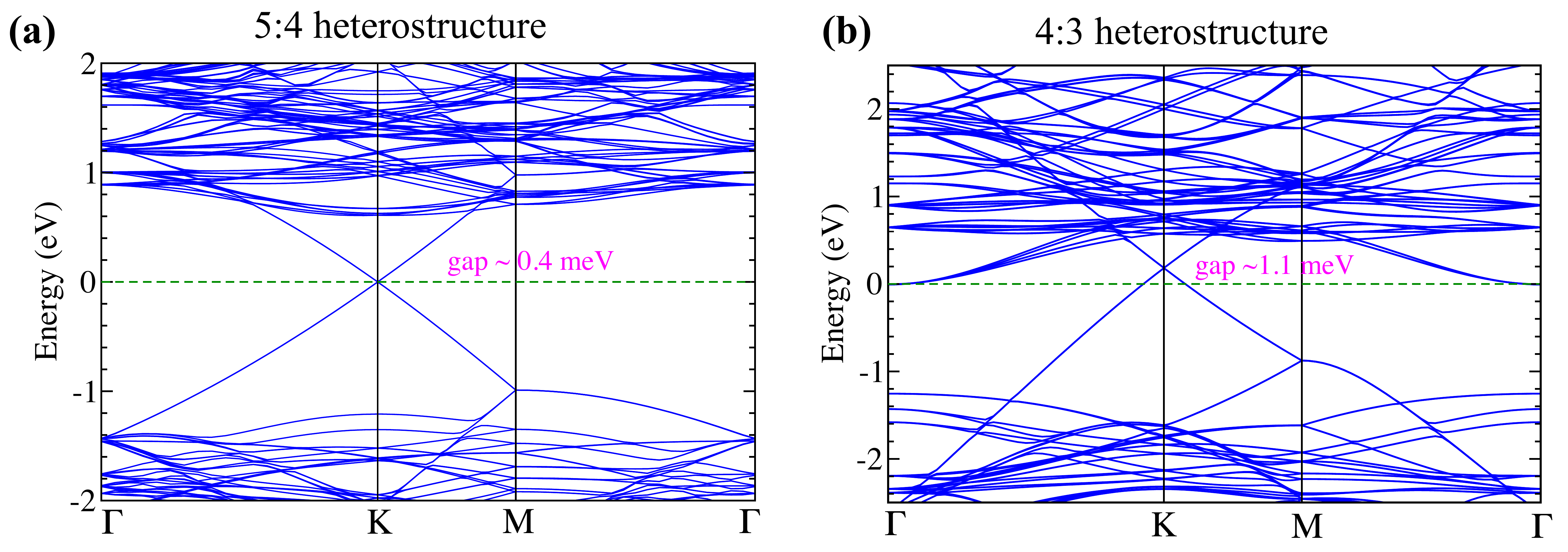}
 \caption{(Color online) Figures (a) and (b) represent the electronic bandstructure of 5:4 and 4:3 graphene/MoS$_{2}$ bilayers calculated using SOC, respectively.}
 \label{fig:bands}
 \end{figure*}

\begin{figure}[hbt!]
 \centering
 \includegraphics[width=8.7cm, keepaspectratio=true]{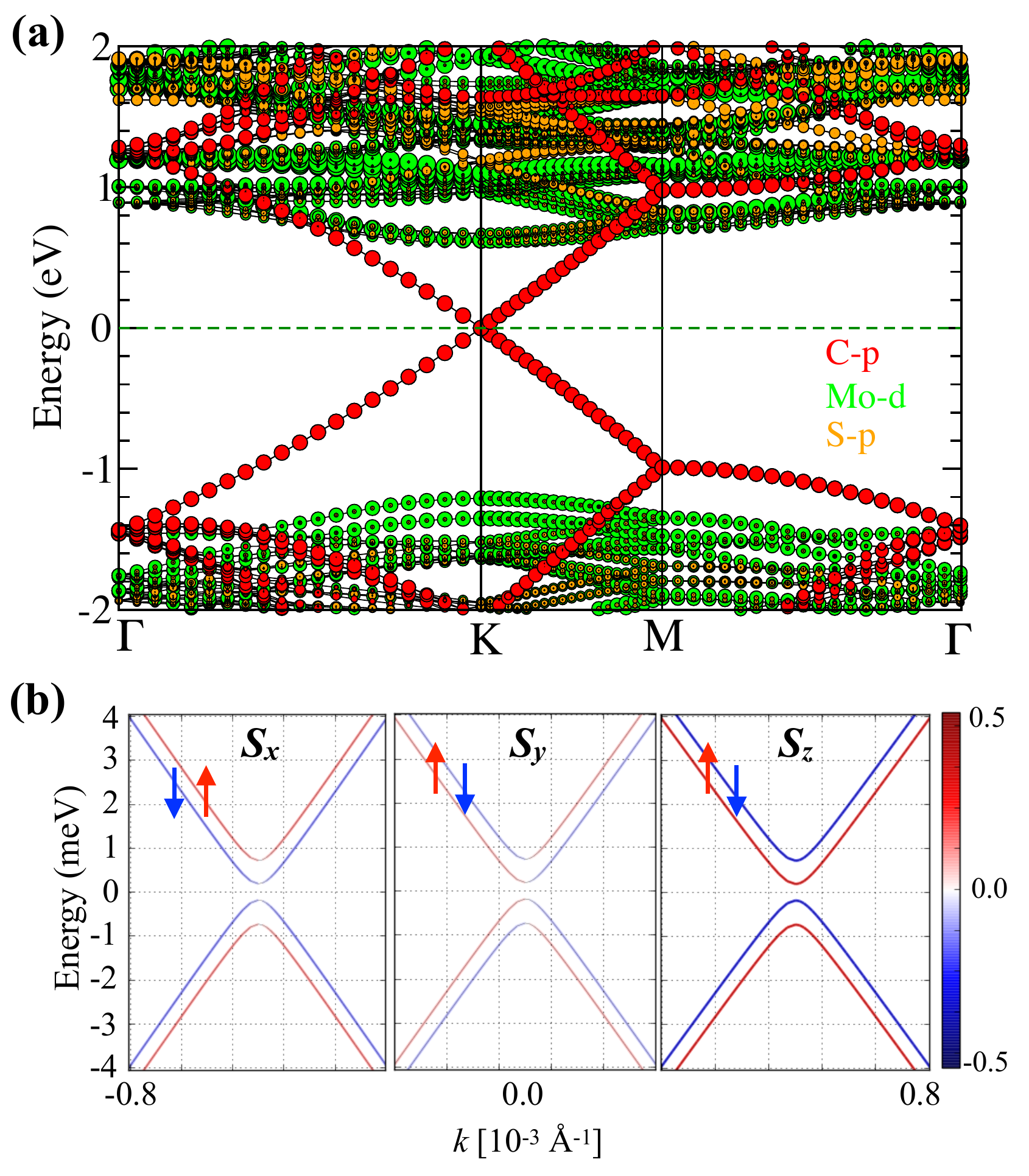}
 \caption{(Color online) Figure shows the orbital projected electronic bandstructure (with-SOC) of 5:4 graphene/MoS$_2$ bilayer heterostructure. Red, green, and orange colors depict projection of C-$p$, Mo-$d$, and S-$p$ orbitals, respectively. (b) Projection of three spin-components on the bands near K point. Here, horizontal axis ranges from [-0.8, 0.8] $\times$ 10$^{-4}$ \AA$^{-1}$ with center defined at K-point.  }    
 \label{fig:orbitals}
 \end{figure}

\subsection{Electronic structure}
Figure~\ref{fig:bands} shows the electronic bandstructure of 5:4 and 4:3 graphene/MoS$_2$ bilayers. The orbital projected electronic bandstructure for 5:4 bilayer heterostructure is given in Fig.~\ref{fig:orbitals}(a). The orbitals character of electronic bands of graphene and MoS$_2$ monolayer are well preserved in 5:4 bilayer heterostructure. In Fig.~\ref{fig:orbitals}, one can notice that the electronic bands forming Dirac cone are coming from C-$2p$ orbitals, whereas the other conduction and valence bands near Fermi-level are mainly composed of Mo-$4d$ and S-$3p$ orbitals. The direct bandgap between the conduction band minima and valence band maxima of MoS$_2$ at K-point is $\sim$1.8 eV, which is in excellent agreement with reported values in the literature. \cite{Ramasubramaniam2012, Espejo2013, Abbas_APL2014, UdoPRB2015, Cecil2017} Dirac point in 5:4 graphene/MoS$_2$ bilayer is located at the Fermi-level and is well-separated from the conduction and valence bands of MoS$_2$, which is in agreement with the experimental observations reporting no charge transfer between the layers at equilibrium conditions.\cite{PierucciNano2016} The conduction and valence bands forming Dirac cone come from the A and B sublattices of graphene, respectively. 

The SOC and proximity effects open a direct bandgap in graphene of $\sim$0.4 meV. In addition to open a direct bandgap at Dirac point, these effects also yield a parabolic shape to the linear bands near K-point. An enlarged view of spin-splitting of bands near K-point is shown in Fig.~\ref{fig:orbitals}(b). Some notable features of the parabolic bands near K-point are: (i) Rashba spin-splitting due to the broken inversion symmetry and SOC effects from MoS$_2$ layer, (ii) opening of a spin gap and anticrossing of bands due to the intrinsic SOC of graphene, and (iii) opening of an orbital gap due to the effective staggered potential arising from the proximity of MoS$_2$ layer. \cite{Gmitra_PRB2009bands, KonschuhPRB2010, gmitra2017proximity} A recent study\cite{Abdulrhman2016} shows that a topological phase transition mediated by band-inversion at K-point can be achieved by utilizing an interlink among the aforementioned competitive interactions. In particular, the competition between the SOC effects from Mo-$d$ orbitals and the intrinsic SOC of graphene when combined with the staggered potential results in topologically distinct regimes where the bilayer heterostructure changes phase from a quantum spin Hall insulator to a normal insulator. In principle, these phases can be controlled by applying a relative gate voltage between the layers.\cite{Abdulrhman2016} Interestingly, in graphene/WSe$_2$ bilayer heterostructure the band-inversion at Dirac point occurs naturally, thanks to the strong SOC of WSe$_2$ layer, indicating presence of a non-trivial topological phase in graphene/WSe$_2$ bilayer. \cite{GmitraPRB2016_WSe2, GmitraPRL2017, gmitra2017proximity}

The direct bandgap at K-point increases almost by three times (from 0.4 meV to 1.1 meV) in the 4:3 graphene/MoS$_2$ bilayer heterostructure. This can be partially attributed to the fact that the 4:3 bilayer inherits larger lattice mismatch compared to that of the 5:4 bilayer, and the interfacial strains often cause variation in the bandgap. Another interesting feature that we observe in 4:3 heterostructure is the charge-transfer between the layers. We notice that the Dirac point of graphene in 4:3 heterostructure is shifted above the lowest conduction band of MoS$_2$, thus indicating donation of electrons to the MoS$_2$ layer by graphene [see Fig.~\ref{fig:bands}(b)]. Also, the valence and conduction bands of MoS$_2$ layer are shifted towards lower energies in 4:3 heterostructure. The tunability of Dirac point and the charge-transfer process between the layers are of central interest for practical applications. \cite{WeiHan2014, RoyNature2013}

Notably, MoS$_2$ undergoes a direct to indirect bandgap transition in 4:3 bilayer heterostructure as can be seen in Fig.~\ref{fig:bands}, while the direct bandgap characteristic of the isolated MoS$_2$ monolayer is preserved in the 5:4 bilayer. Change from direct to indirect bandgap is observed even with just two layers of MoS$_2$ without introducing any strain effects. \cite{MoS2Galli2, NaikPRB2017} A recent work investigated the physical origin of layer dependence in bandstructure of two-dimensional materials, and concluded that in addition to the quantum confinement effects, the nonlinearity of exchange-correlation functional plays a crucial role in determining the direct to indirect bandgap transition in two-dimensional materials.\cite{NaikPRB2017} In case of the graphene/MoS$_2$ bilayer having minimal strain i.e. the 5:4 bilayer, the gap transition is not present which indicates that the electronic interactions between graphene and MoS$_2$ layer in 5:4 bilayer are subtler than that of in pristine MoS$_2$ bilayer, and hence the transition in the 4:3 bilayer is triggered by the imposed strain to the layers. This finding is important since it suggests that the MoS$_2$ monolayer supported on graphene can present high photoluminescence as it was previously found on samples supported on SiO$_2$ \cite{MoS2Galli2}, and described theoretically in reference.\cite{MoS2onSiO2} The electronic bandstructures calculated for graphene/MoS$_2$ bilayers using {\sc ABINIT}\cite{GONZE2016106} code further corroborate this finding. 

Based on the comparative analysis of 5:4 and 4:3 bilayer heterostructures, we argue that the existing controversy regarding the electronic bandstructure could be due to the inadequate evaluation of weak non-local vdW effects in these semi-metallic/semi-insulating heterostructures, and different lattice mismatches that were considered in the independently reported studies. Since the effect of biaxial strain, stacking order, and interlayer twist on the electronic structure of bilayer heterostructures has already been reported in the literature, \cite{XDLiACS2013, CDiaz2015, Zilu2015, oakes2016interface, Tribhuwan2016, LXiaolong2016, LXingen2015, LiangXu2017} in this work we investigate the effect of uniaxial stress along $c$-axis on the electronic bandstructure of graphene/MoS$_2$ bilayer heterostructures.

\begin{figure}
\includegraphics[width=8.5cm]{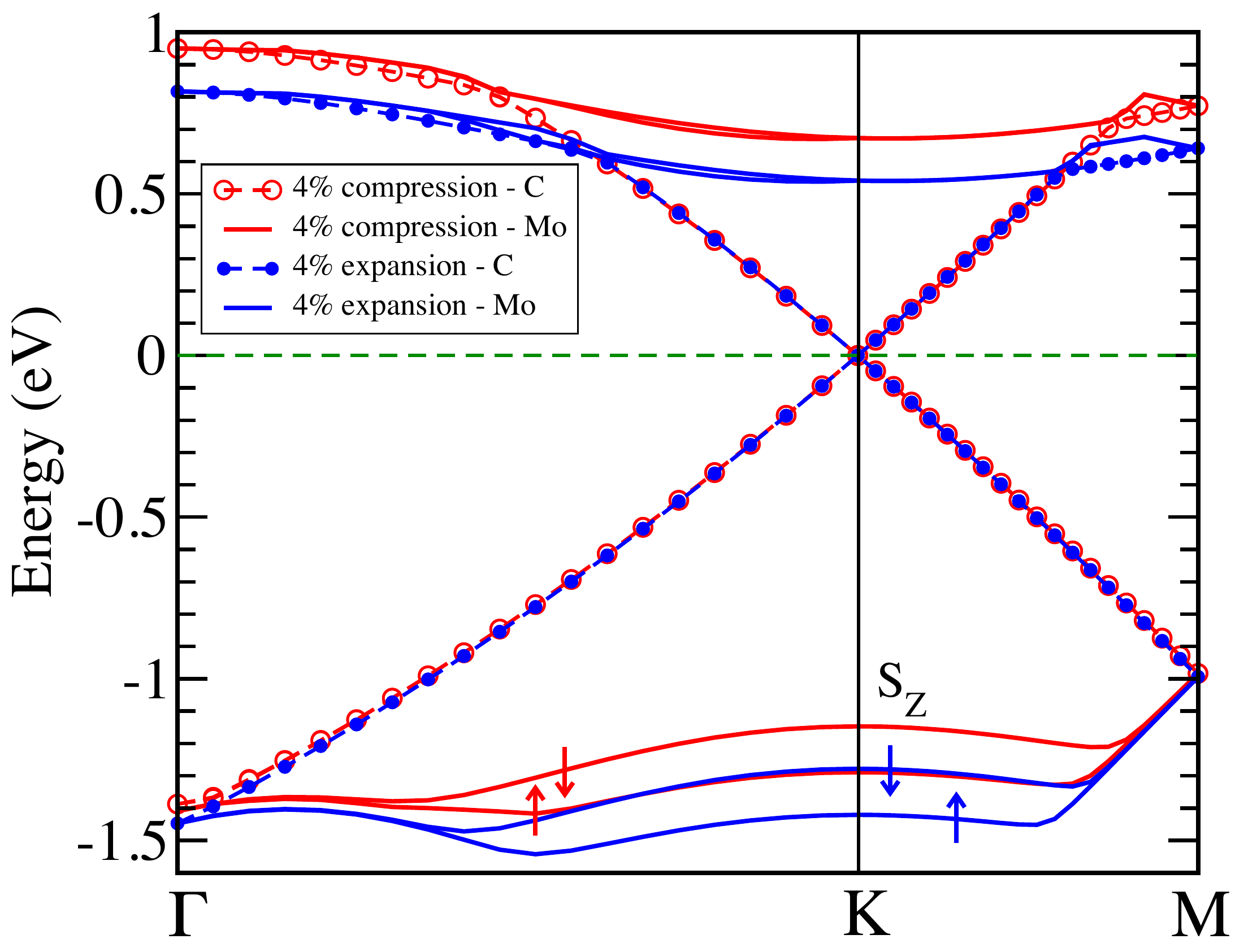}
\caption{(Color online) SOC calculated electronic bandstructure of graphene/MoS$_2$ bilayer at 4\% compression (red) and 4\% expansion (blue) near the Fermi level. Circles depict the bands from C atoms while solid lines depict bands from Mo atoms. The direction of arrows shows the $S_Z$ spin-component. }
\label{dist-bands}
\end{figure}

{\it Effect of interlayer spacing on the electronic bandstructure}: The effect of interlayer spacing on the electronic structure of the 5:4 bilayer heterostructure was studied by performing electronic bandstructure calculations at four values of interlayer strains, namely $x$ = -4\%, -2\%, +2\%, and +4\%, where positive (negative) values refer to expansion (compression) from the equilibrium interlayer distance. Atoms were allowed to relax in the plane of layers while their vertical coordinates were kept fixed at each separation. Figure~\ref{dist-bands} displays the bands (calculated with SOC) corresponding to the four upper valence bands and the four lower conduction bands near Fermi-level, for separations corresponding to 4\% expansion (blue) and 4\% compression (red) with respect to the equilibrium interlayer distance. The effect of diminishing interlayer distance is that the Mo bands shift to higher energy values with respect to the Fermi level. Since the increment is same for conduction and valence bands, the direct bandgap of MoS$_2$ monolayer is not modified, as can be seen in Fig.~\ref{dist-bands}. At $K$ point of Brillouin zone there is a spin splitting of Mo valence bands, which is depicted by arrows in Fig.~\ref{dist-bands}. Such spin splitting is also present in the Mo conduction bands, however at $K$ point the conduction bands maintain the spin-degeneracy. The spin splitting of Mo valence bands is $\sim$0.2 eV, which remains constant in the range of studied interlayer separations.  

Regarding variation in the direct bandgap at Dirac point, we find that the gap at Dirac point increases substantially from $\sim$0.2 meV to $\sim$0.7 meV when going from 4\% expansion to 4\% compression. However, other features of bands were preserved in the case of uniaxial strains. At larger compression, i.e. at $x$ = -4\%, the $p_z$-orbitals of graphene strongly interact with Mo-$d$ orbitals, thus resulting larger bandgap at Dirac point. However, with increasing interlayer separation ($x$), the direct bandgap at Dirac point is expected to decrease systematically and attain the value for isolated graphene. On the other hand, the direct bandgap of MoS$_2$ (at $K$) is robust enough to be unaffected by the proximity of graphene even at 4\% compression. Nonetheless, the direct bandgap at $\Gamma$ systematically changes from 2.26 eV to 2.34 eV with increasing interlayer compression from $x$ = +4\% to $x$ = -4\%. 

\begin{figure}
 \includegraphics[width=8.8cm, keepaspectratio=true]{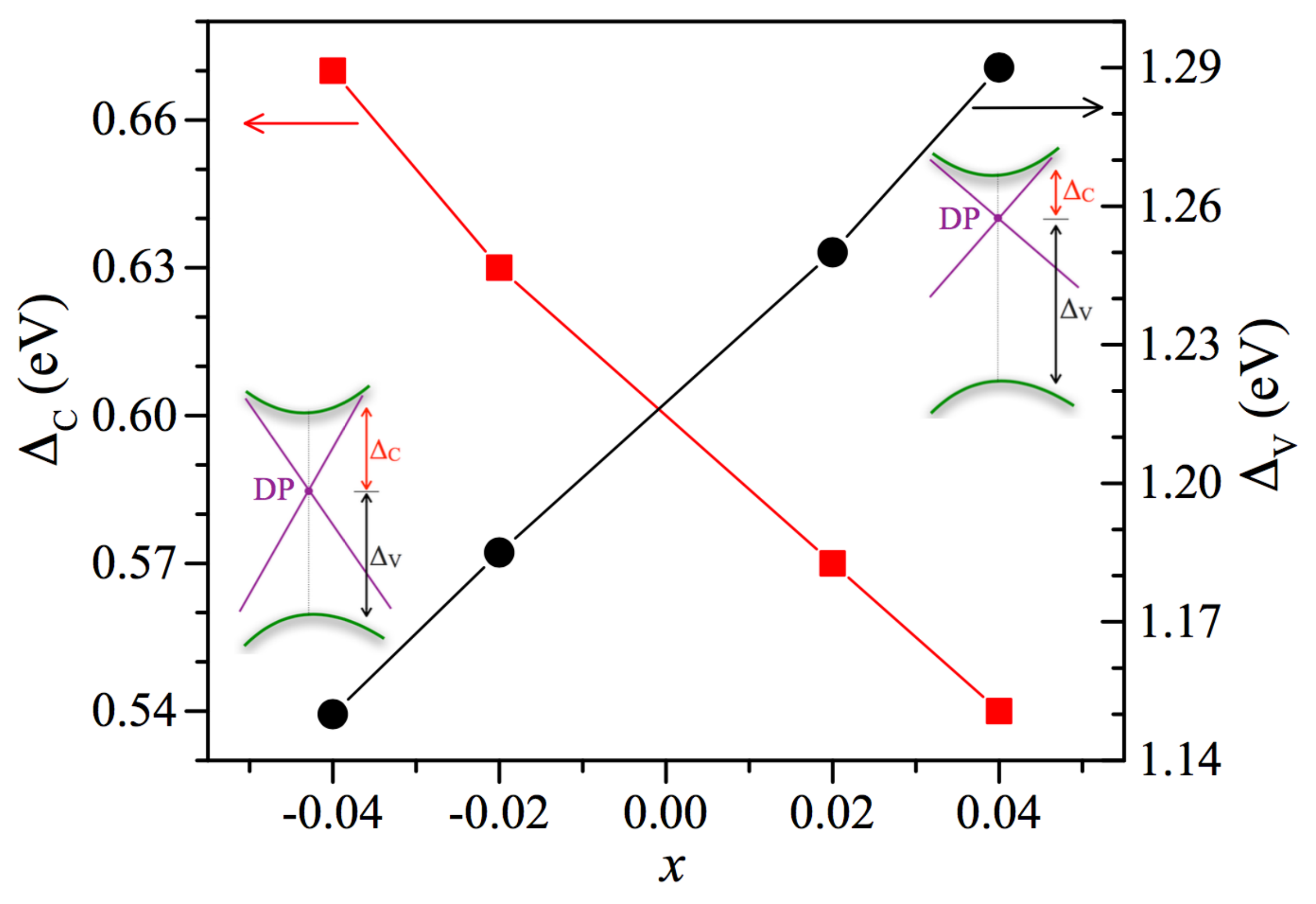}
\caption{(Color online) Figure shows change in the location of Dirac point for different values of interlayer separations in 5:4 graphene/MoS$_2$ bilayer. $\Delta_{C}$ ($\Delta_{V}$) represents the magnitude of energy separation between the Dirac point and the lowest conduction (valence) band of MoS$_2$ layer, as illustrated in the inset. Green lines depict the bands from MoS$_2$ near $K$ point. $x$ is the magnitude of the uniaxial strain on the interlayer separation. The positive (negative) values of $x$ refer to expansion (compression). }
\label{shift_DP}
\end{figure}

Fig.~\ref{shift_DP} illustrates change in the location of Dirac point due to the varying interlayer separation between graphene and MoS$_2$ nanosheets. We observe that Dirac point shifts towards the MoS$_2$ valence bands (i.e. increasing $\Delta_{C}$) as we increase the compressive interlayer strain from +4\% to -4\%. This can be ascribed to the enhanced electric field effects arising due to the asymmetric interlayer potential. Such out-of-plane electric field or gate bias effects have also been demonstrated to open a gap at Dirac point in graphene. \cite{McCannPRB2006, CastroPRL2007, MakPRL2009, Koshino2010, Suarez2011, Munoz2016}

\begin{figure*}
 \includegraphics[width=18cm, keepaspectratio=true]{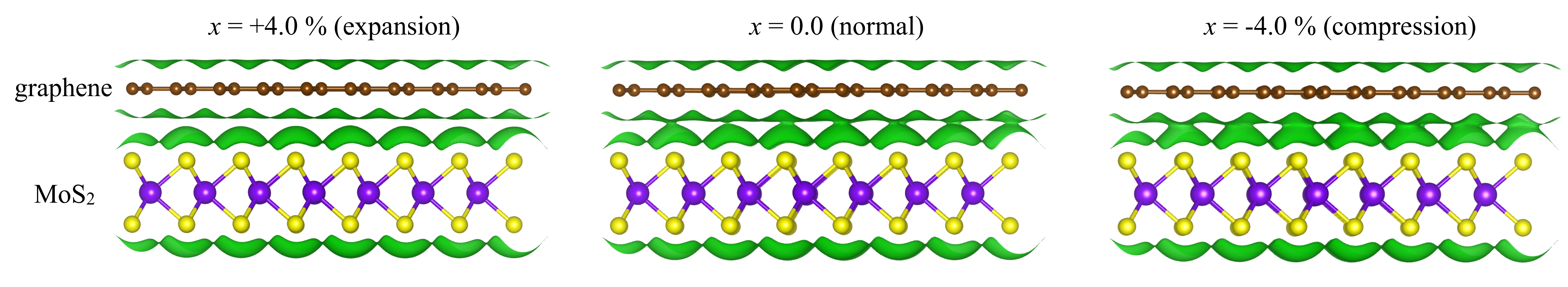}
\caption{(Color online) Isodensity surfaces (green color) at isosurface level $n$=0.007 a.u. for graphene/MoS$_2$ bilayer heterostructures: (left) 4\% interlayer expansion, (center) optimized interlayer separation, and (right) 4\% interlayer compression from equilibrium.}
\label{isosurf}
\end{figure*}

In order to appreciate any possible changes in the electronic density at the interface, charge isosurface has been plotted in Fig.~\ref{isosurf}. Differences from the isolated layers case can be detected for low values of electronic densities, in fact an overlap between the charge densities of both monolayers is found for $n$=0.007 a.u. at ambient condition ($x$ = 0), which increases with increasing interlayer compression, however, such charge density overlap is not present for the 4\% expansion, as depicted in Fig.~\ref{isosurf}. Such charge accumulation between the graphene and MoS$_2$ nanosheets aids to the opening of bandgap at Dirac point, as suggested by McCann. \cite{McCannPRB2006} Although uniaxial strains effects were found to cause notable changes in the electronic properties of graphene/MoS$_2$ bilayers, no significant changes were observed in the Mo-S, S-S and C-C bond lengths due to the varying interlayer separation in the studied range.

\begin{figure*}[hbt!]
 \centering
 \includegraphics[width=18cm, keepaspectratio=true]{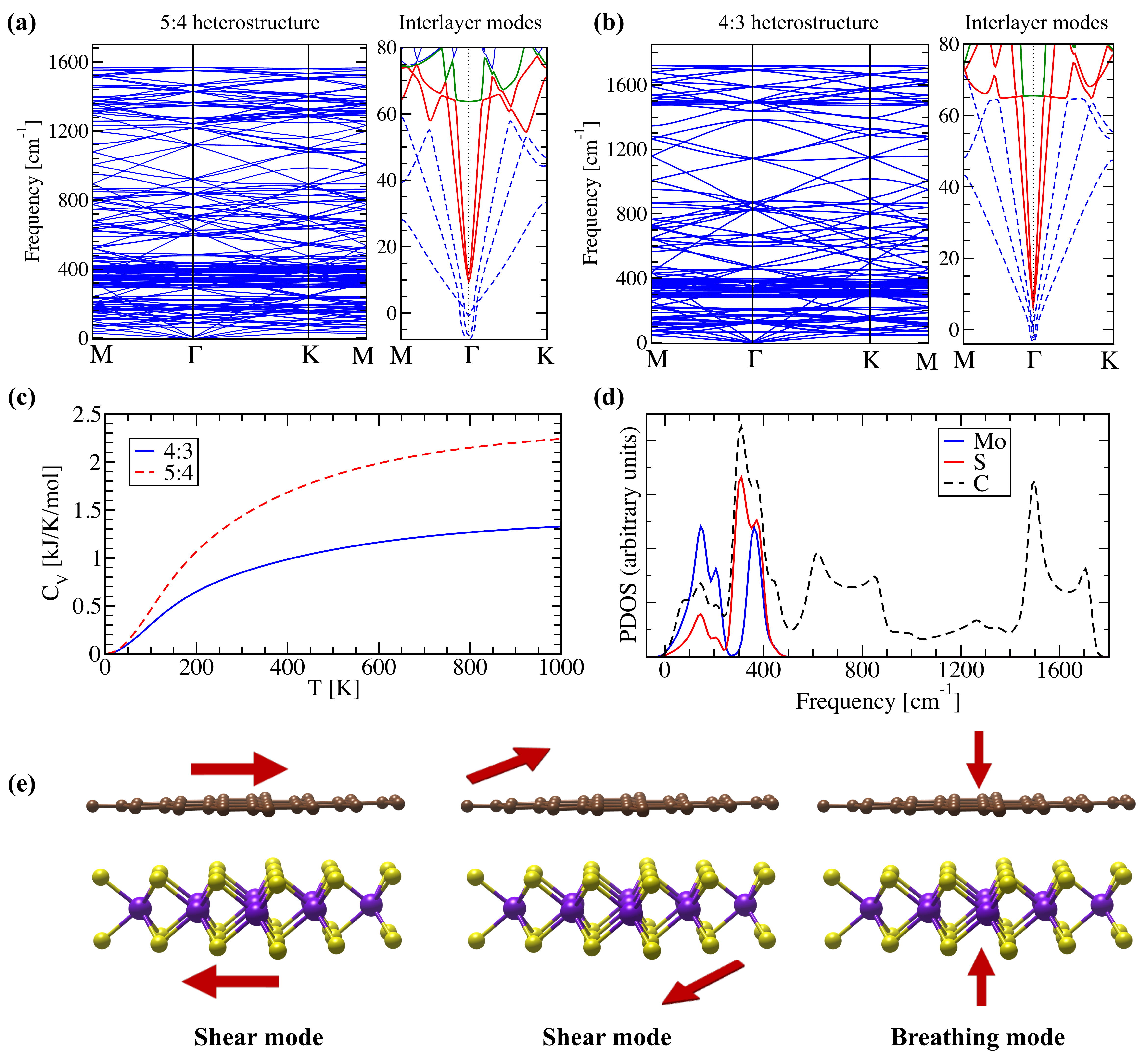}
 \caption{(Color online) Figures (a) and (b) show the calculated phonon dispersion of 5:4 and 4:3 graphene/MoS$_2$ bilayer heterostructures, respectively. An enlarged view of the interlayer phonon modes, i.e. two shear modes (red solid lines) and one breathing mode (green solid line), is depicted next to each phonon spectra. Three dotted blue lines represent three acoustic phonon modes ($ZA, TA, LA$). (c) Calculated specific heat. (d) Atom projected phonon density of states (PDOS)for 4:3 heterostructure. A similar spectra with slight variation in frequencies was obtained for 5:4 heterostructure. (e) A schematic illustration of longitudinal shear, transverse shear and breathing interlayer phonon modes in graphene/MoS$_2$ bilayer.}
  \label{fig:phonons}
 \end{figure*}

 \begin{table}[hbt!]
\centering
\caption{Group velocities ($\nu$) for transverse acoustic ($TA$), longitudinal acoustic ($LA$), transverse shear ($S_{T}$), and longitudinal shear ($S_{L}$) phonon branches. \\ }
\begin{adjustbox}{width=0.48\textwidth} 
\begin{tabular}{L c c c c}
  \hline
Structure & $\nu_{TA}$ ($m/s$) &  $\nu_{LA}$ ($m/s$) &  $S_{T} $ ($m/s$) & $S_{L}$ ($m/s$) \\
\hline\\[-2mm]
5:4 heterostructure  &  3877 &  6626  &  12484  &   21801 \\
4:3 heterostructure  &  4262 &  6582  &  12692  &   20319 \\

\hline
\end{tabular}
\end{adjustbox}
\label{table2}
\end{table}

\subsection{Vibrational properties of graphene/MoS$_2$ bilayer heterostructures}
A thorough understanding of the vibrational properties of vdW heterostructures is not only important for fundamental physics, but it also renders great insights into the observed mechanical, thermal and electronic properties in vdW heterostructures. Several recent studies have primarily focused on the vibrational properties of graphene\cite{Yan2008, Wang2009, Lui2013}, graphite\cite{WIRTZ2004}, transition metal dichalcogenides (TMDs) \cite{DING2011, Zhao2013}, and their stacked vdW heterostructures\cite{Lui2015, NamLe2016}. The interlayer phonon modes that arise due to the weak vdW interactions are of special interest in vdW structures. \cite{Lui2013, Zhao2013, Lui2015, NamLe2016} Therefore, in this section we investigate the vibrational properties of graphene/MoS$_2$ 2D heterostructures with a special focus on the observed interlayer phonon modes in the studied systems. 

Figure~\ref{fig:phonons} shows the phonon spectra, total and atom projected phonon density of states (PDOS) for 5:4 and 4:3 bilayer graphene/MoS$_2$ structures, calculated using the {\sc DFPT} approach as implemented in the {\sc VASP} code. The SOC and vdW were included in the phonon calculations. We find that all the phonon modes have positive frequency (except for a very small imaginary frequency near $\Gamma$ which are often present in the theoretical calculations for 2D systems due to the inadequate numerical convergence close to $k$=0~\cite{Sobhit2017}). The positive phonon frequencies guarantee the thermodynamical stability of the bilayer structures. The three lowest frequency phonon branches constitute $ZA$, $TA$ and $LA$ acoustic modes. We also observe features of the flexural acoustic mode in the low frequency phonon dispersion very near the $\Gamma$ point. Flexural modes are particularly important for 2D systems and have been widely studied in 2D layered structures. \cite{Jiang2015} These modes significantly  contribute  to the PDOS and are responsible for the large thermal conductance. Le et al. \cite{NamLe2016} have reported that the transport due to the flexural excitations is almost ballistic. 

Next to the three acoustic modes there exists the interlayer shear and breathing phonon modes. These vibrational modes describe the relative displacement of graphene sheet with respect to the MoS$_2$ sheet in tangential (in-plane) and perpendicular (out-of-plane) directions. From the crystal symmetry, we expect two shear phonon modes and one breathing phonon mode as illustrated in Fig.~\ref{fig:phonons}(e). The breathing phonon mode is of particular importance since it is very sensitive to the interlayer vdW interaction and its count $(N-1)$ increases with increasing number of layers $(N$) in a multilayer 2D vdW heterostructure. Depending upon $N$, these low frequency modes can be Raman-active and/or IR-active.\cite{Zhao2013} In 5:4 (4:3) graphene/MoS$_2$ bilayer, the frequency of the two shear modes and one breathing phonon mode at $\Gamma$ point is 9.53 (6.45), 11.75 (8.69) and 63.75 (65.54) cm$^{-1}$, respectively. Note that the frequency of the breathing phonon mode is almost 10 times larger than that of the shear phonon modes. Moreover, the breathing phonon mode has almost zero dispersion in the phonon spectra near $\Gamma$. These interlayer phonons modes play a vital role in understanding different underlying scattering mechanisms in layered 2D structures, and are very sensitive to the vDW interactions. The characteristic group velocity of the transverse acoustic ($TA$), longitudinal acoustic ($LA$) and interlayer shear phonon modes is given Table 2. Our results are in good agreement with a recent work reported by Le al. \cite{NamLe2016} The atom resolved PDOS spectra reveals that all Mo, S and C atoms contribute to the acoustic modes, whereas, high frequency optical modes ($ > $ 500 $cm^{-1}$) have contribution only from the C atoms [see Fig.~\ref{fig:phonons}(d)]. Although the acoustic phonons mainly govern the heat transfer process, the optical phonons provide various scattering channels which could reduce the acoustic phonon life-times through acoustic-optic phonon scattering mechanisms.\cite{PAVLIC201715} A very low lattice thermal conductivity is expected in graphene/MoS$_2$ heterostructures along the vertical stacking direction due to the weak vdW bonding. The presence of multiple phonon band crossings in the phonon spectra (see Fig.~\ref{fig:phonons}) calls for a detailed investigation of the lattice thermal conductivity in graphene/MoS$_2$ heterostructures. 

The specific heat (C$_V$) was determined from the phonons for both 5:4 and 4:3 bilayer heterostructures, and is shown in Fig.~\ref{fig:phonons}(c). Graphene/MoS$_2$ bilayer heterostructures exhibit very large heat capacity compared to that of the constituent monolayers.\cite{ToheiPRB2006, YUAN20151} Due to the large Debye temperature of graphene ($\sim$1000 K)\cite{ToheiPRB2006}, the heat capacity of heterostructures approaches to its Dulong-Petit limit at high temperatures near $\sim$1000 K. Below 1000 K it follows the $T^3$ power law due to the dominant contribution from the lattice vibrations. The higher phonon density of states in 5:4 bilayer can be accounted for its larger heat capacity compared to that of in 4:3 bilayer. Due to their high heat capacity and outstanding cycling stability, graphene/MoS$_2$ heterostructures are promising anode materials for batteries. \cite{Chang2011, ChangCC2011, ShaoJPCC2015, fu2014situ}

\section{Elastic Properties}
The knowledge of elastic properties not only provides deep insights in understanding the nature of vdW interactions in graphene/MoS$_2$ vdW heterostructures, but it is also essential for practical applications of such heterostructures in modern technology. Therefore, we have systematically evaluated the elastic stiffness constants ($C_{ij}$) for 4:3 and 5:4 bilayer heterostructures using the stress-strain relationship as implemented in {\sc VASP} code \cite{Kresse1996, Kresse1999}, and further determined various elastic moduli using {\sc MechElastic} script \cite{MechElastic, singh2017elastic}. The $C_{ij}$ values of heterostructures are converged better than 2.0 N/m by increasing the $k$-mesh size. Due to the hexagonal crystal geometry of system, only $C_{11} (= C_{22}$), $C_{12}$, and $C_{66}$ values are relevant. The 2D layer modulus, a quantity that represents the resistance of a nanosheet to stretching, can be calculated for hexagonal systems using equation:\cite{AndrewPRB_2012} 

\begin{equation}
\gamma^{2D} = \frac{1}{2}[C_{11} + C_{12}]
\end{equation}

 \begin{table*}[hbt!]
\centering
\caption{List of the calculated elastic moduli (in N/m units) for graphene/MoS$_2$ bilayer heterostructures, isolated graphene and MoS$_{2}$ monolayer.\\ }

\begin{adjustbox}{width=1.0\textwidth} 
\setlength{\tabcolsep}{8pt}
   \setlength{\arrayrulewidth}{0.3mm}
\renewcommand{\arraystretch}{1.5}
     \begin{threeparttable}
\begin{tabular}{c  c c c c c c c}
 \hline
             &  $C_{11}$  & $C_{12}$ & Shear modulus (=$C_{66}$ ) &  Layer modulus &  Young's modulus &  Poisson's ratio &   \\
 \hline
 graphene/MoS$_2$ bilayer (4:3) & 500.0  & 109.7   & 194.0   & 304.2    & 395.0  & 0.22 & This work \\
 graphene/MoS$_2$ bilayer (5:4) & 492.4  & 83.6   & 203.2   & 286.4    & 427.0  & 0.17 & This work\\
 graphene/MoS$_2$ bilayer  & -  & -   & -   & -    & 467$\pm$48  & - & Ref.\cite{KaiNano2014} (Exp.)\\
 \hline
 \multirow{5}{*}{graphene} & 358.9  & 65.1   & 146.9  & 212.0  & 347.1  & 0.18 & This work \tnote{a} \\	
     					   & 352.7  & 60.9   & 145.9  & 206.6  & 342.2  & 0.17 & Ref.\cite{AndrewPRB_2012} (Theory) \\
     					   & 358.1  & 60.4   & 148.9\tnote{b}  & 209.3\tnote{b}  & 348.0  & 0.17 & Ref.\cite{Xiaoding2009} (Theory) \\
     					   & -  & -   & -  & -  & 342$\pm$40  & 0.17 & Ref.\cite{Lee385} (Exp.) \\
     					   & -  & -   & -  & -  & 349$\pm$12  & - & Ref.\cite{KaiNano2014} (Exp.) \\

\hline					
 \multirow{4}{*}{MoS$_2$} & 132.3 & 32.8  & 49.5  & 82.5  & 124.1  & 0.25 & This work\tnote{a} \\	
 					 & 128.4 & 32.6 &    47.9\tnote{b}   &  80.5\tnote{b}  &  120.1 &  0.25\tnote{b}  & Ref.\cite{Qing_MoS2-2013} (Theory) \\
					 & 130 & 40 & 45\tnote{b}  &  85\tnote{b}   &  117.7\tnote{b}  &  0.29  & Ref.\cite{Cooper2013} (Theory) \\
					 & - & - & -  &  -   &  123  &  0.25  & Ref.\cite{KaiNano2014} (Exp.) \\

\hline \\
\end{tabular}

    \begin{tablenotes}
        \item[a] Elastic properties of isolated monolayer were calculated using {\sc VASP}(PBE), and converged to less than 1.0 N/m by increasing $k$-mesh size. 
        \item[b] Estimated from the data reported in the reference. 
    \end{tablenotes}
 \end{threeparttable}

\end{adjustbox}
\label{table3}
\end{table*}

Average 2D Young's modulus ($E$), Poisson's ratio ($\nu$), and shear modulus ($G$) can be obtained using following expressions: 

\begin{equation}
\begin{split}
E = \frac{C_{11}^{2} - C_{12}^{2}}{C_{11}}, \\ \\
\nu = C_{12}/C_{11}, \\ \\
G = C_{66}.
\end{split}
\end{equation}

Table 3 shows a list of the calculated elastic moduli for graphene/MoS$_2$ bilayer heterostructures, isolated graphene and MoS$_{2}$ monolayers. Our data are in excellent agreement with the reported values in literature from theoretical and experimental studies. We notice that although isolated graphene and MoS$_2$ monolayer inherit complimentary physical properties, their combination mitigates the adverse elastic properties of each individual constituent providing a novel platform to engineer their properties. We notice that the $C_{ij}$ values for graphene/MoS$_2$ bilayers are roughly arithmetic sum of the $C_{ij}$ values for isolated graphene and  MoS$_2$ monolayers. Overall elastic properties of bilayer heterostructure are better compared to that of the constituent monolayers. Notably, the elastic stiffness constants (except $C_{66}$) attain lower values ($i.e.$ elastic softening) in 5:4 bilayer compared to that of in 4:3 bilayer. The following two reasons are primarily responsible for the observed elastic softening: (i) The 5:4 bilayer inherits lower lattice mismatch than 4:3 bilayer, therefore it suffers lesser in-plane strain energy, and thereby reduces the effective elastic stiffness, and (ii) the nonlinear elastic response of the constituent monolayers.\cite{Lee385, Cooper2013} The coefficient of the second-order term in the nonlinear elastic response is generally negative for most of the materials, which leads to a decrease (increase) in the elastic stiffness at large tensile (compressive) strains.\cite{Lee385} In 5:4 bilayer, the graphene sheet undergoes a tensile strain of $\sim$1.2\%, whereas graphene sheet undergoes a compressive strain of $\sim$1.4\% in 4:3 bilayer heterostructure. On the other hand, the larger interlayer spacing and hence lesser vdW energy ($|0.15|$ eV/atom) in 4:3 bilayer could be held accountable for lower shear elastic modulus compared to that of in 5:4 bilayer (vdW energy $\approx |0.17|$ eV/atom). 

We further study the intrinsic strength, bending modulus, and buckling phenomenon in graphene/MoS$_2$ heterostructures. The intrinsic strength ($\sigma_{int}$) can be estimated using the Griffith's proposal: $\sigma_{int}$ $\sim$ $\frac{E}{9}$.\cite{Griffith163, Cooper2013} From values listed in Table 3, one can notice that the intrinsic strength of bilayer heterostructures is considerably larger than that of the isolated monolayers. The bending modulus ($D$) for a 2D nanosheet can be calculated using equation:\cite{Jiang_review2015}

\begin{equation}
D = \frac{E h^{2}}{12(1-\nu^{2})},
\end{equation}

where, $h$ is the thickness of the nanosheet. Accurate determination of $h$ is uncertain because of the electronic configuration along the normal direction, which is subject to change under deformation. Due to this uncertinity, $D$ could acquire different values depending upon the chosen $h$. However, the lower estimate of $D$ can be calculated using the absolute thickness of the nanosheet. For example: the absolute thickness of graphene is 0.6 -- 0.8 \AA\cite{TuPRB2002}, and for MoS$_2$ is $\sim$3.13 \AA~(see Fig.~\ref{fig:mono_crystal-phonons}). Considering the absolute thickness of graphene/MoS$_2$ 5:4 bilayer heterostructure ($h$ = 3.13+3.40+0.75\cite{TuPRB2002} = 7.28 \AA), the obtained $D$ is 121.2 eV. The experimental value of $h$ for graphene/MoS$_2$ heterostructures lies in range 10 -- 11 \AA.\cite{LiliACSNano2014, Zhang_gMoS2-2014,fu2014situ, Miwa2015} For $h$ = 10.5 \AA, obtained $D$ = 252.2 eV. The reported $D$ values (lower estimates) for graphene and MoS$_2$ monolayers are $\sim$1.2 eV and 9.61 eV.\cite{Jiang_review2015} It is worth to note that although it is easier to bend an isolated graphene and MoS$_2$ monolayer, their combination yields very large bending energy due to its multilayer atomic structure, which offers more interaction terms restraining the bending motion. 

From the knowledge of quantities $D$ and $E$, one can study the buckling phenomenon and estimate the critical buckling strain ($\epsilon_c$) using the Euler's buckling theorem:\cite{Jiang_review2015} 

\begin{equation}
\epsilon_c = -\frac{4\pi^{2} D}{E L^{2}},
\end{equation}

where, $L$ is the length of 2D nanosheet. For isolated graphene, MoS$_2$ monolayer, and graphene/MoS$_2$ 5:4 bilayer heterostructure, we obtain following expressions: 

\begin{center}
\begin{equation}
\begin{split}
{\epsilon_c}^{graphene} = -\frac{2.2}{L^{2}}, \\ \\ 
{\epsilon_c}^{MoS_{2}} = -\frac{48.9}{L^{2}}, \\ \\
{\epsilon_c}^{graphene/MoS_{2}} = -\frac{179.2}{L^{2}}. \\
\end{split}
\end{equation}
\end{center}

Here, $L$ is in \AA~units. For the same length samples, the critical buckling strain for graphene/MoS$_2$ 5:4 bilayer heterostructure (MoS$_2$ monolayer) is more than eighty (twenty) times larger compared to that of in graphene. Therefore, graphene/MoS$_2$ bilayer heterostructures are more robust for in-plane structural deformations and do not buckle easily compared to the individual constituent layers. Hence, in this respect, graphene/MoS$_2$ bilayer heterostructures are better fit for practical applications.

\section{Conclusions}
In this study we report that compared to the available DFT vdW methods, the DFT-TS method best describes the weak vdW interactions and predicts the interlayer spacing accurately in graphene/MoS$_2$ vdW heterostructures. The key reason behind the success of this method is the fact that the local variations of the charge-density near the interface are well captured in the TS-method. Therefore, the TS-method appears to be the best candidate to evaluate the weak vdW interactions at (semi)metallic/insulating interfaces. The predicted interlayer spacing for graphene/MoS$_2$  bilayer structure (5$\times$5 / 4$\times$4) is 3.40 \AA~ which is in excellent agreement with the experimental data. The electronic bandstructure analysis of 5:4 and 4:3 graphene/MoS$_2$ bilayers reveals that the Dirac point of graphene is shifted upwards above the Fermi-level and is located near the conduction bands of MoS$_2$ sheet, yielding a considerable charge-transfer process in 4:3 bilayer, whereas the Dirac point lies in the bandgap region in 5:4 bilayer indicating no charge-transfer process. We find that the location of Dirac point can be shifted by tuning the interlayer spacing between the graphene and MoS$_2$ sheets. The vibrational spectra of 5:4 and 4:3 graphene/MoS$_2$ bilayers reveals the presence of interlayer shear and breathing phonon modes in the bilayers. These interlayer phonons modes play a vital role in understanding different underlying scattering mechanisms in layered 2D structures. The graphene/MoS$_2$ bilayer heterostructures possess large heat capacity, and exhibit much better elastic and mechanical properties compared to that of the isolated constituent monolayers. Elastic stiffness constants, elastic moduli, intrinsic strength, bending modulus, and buckling phenomenon for isolated graphene, MoS$_2$ monolayer, and graphene/MoS$_2$ bilayer heterostructures have been discussed along with a comparison with the available data in the literature. \\


\textit{Acknowledgments}: This work used the Extreme Science and Engineering Discovery Environment (XSEDE), which is supported by National Science Foundation grant number OCI-1053575. Additionally, the authors acknowledge the support from Texas Advances Computer Center (TACC), Bridges supercomputer at Pittsburgh Supercomputer Center and Super Computing Systems (Spruce and Mountaineer) at West Virginia University (WVU). A.H.R. and S.S. acknowledge the support from National Science Foundation (NSF) DMREF-NSF 1434897 and DOE DE-SC0016176 projects. S.S. thanks the donors of Jefimenko family for their financial support through the Oleg D. and Valentina P. Jefimenko Physics Fellowship at WVU. S.S. also acknowledges the support from the Robert T. Bruhn research award and the WVU Foundation Distinguished Doctoral Fellowship. C.E. acknowledges support from Direcci\'on de Investigaci\'on, Creación y Extensi\'on at UTADEO.  Authors thank Prof. Sergio E. Ulloa for fruitful discussions.

\bibliography{./graphene_MoS2_2D_hetero}

\end{document}